\documentstyle[graphicx,aps,multicol]{revtex}
\draft

\begin{document}

\title{Thermal and electromagnetic properties of $^{166}$Er and $^{167}$Er}

\author{E.~Melby, M.~Guttormsen, J.~Rekstad, A.~Schiller, and S.~Siem}
\address{Department of Physics, University of Oslo, N-0316 Oslo, Norway}
\author{A.~Voinov}
\address{Frank Laboratory of Neutron Physics, Joint Institute of Nuclear
Research, 141980 Dubna, Moscow reg., Russia}
\maketitle

\begin{abstract}
The primary $\gamma$-ray spectra of $^{166}$Er and
$^{167}$Er are deduced from the ($^3$He,$\alpha \gamma$) and
($^3$He,$^3$He'$\gamma$) reaction, respectively, enabling a simultaneous extraction of the level density and the $\gamma$-ray strength function.
Entropy, temperature and heat capacity are deduced from the level density within the micro-canonical and the canonical ensemble, displaying signals of a phase-like transition from the pair-correlated ground state to an uncorrelated state at $T_c\sim 0.5$ MeV.
The $\gamma$-ray strength function displays a bump around $E_{\gamma}\sim 3$ MeV, interpreted as the pygmy resonance.
\end{abstract}


\begin{multicols}{2}

\section{Introduction}

The energy distribution of primary $\gamma$-rays provides information on both the level density and the $\gamma$-ray strength function.
The present aim is to study the level density and $\gamma$-strength function of the nuclei $^{166,167}$Er and to compare with other nuclei in this mass region.
In addition, thermodynamic and electromagnetic properties of the two nuclei will be investigated. 

Most of the experimental information on the level density of rare earth nuclei originates from the neutron-resonance spacing at the neutron-separation energy~\cite{ilj92} and direct counting of levels in the vicinity of the ground state~\cite{toi}.
Experimental knowledge of the $\gamma$-strength function is mainly based on the study of photo-absorption cross sections~\cite{AogN} and radiative neutron capture~\cite{becvar}.

A new method~\cite{andreas:nim} derives the level density and $\gamma$-strength function simultaneously from a set of primary $\gamma$ spectra.
This technique has proven to give a valuable supplement to the more traditional methods for level density extraction, and to our knowledge, it represents the least model-dependent method to obtain the $\gamma$ strength function over a wide energy region below the neutron-separation energy.

The nuclear level density is closely related to the thermodynamic properties of nuclei, and can therefore be utilized to deduce e.g.~entropy, temperature and heat capacity of nuclei.
The thermodynamic observables may reveal phase transitions.

The $\gamma$-strength function is a measure for the average electromagnetic properties of nuclei and has a fundamental importance for the understanding of nuclear structure and reactions involving $\gamma$ rays~\cite{kop}.
At $\sim 3$ MeV of $\gamma$ energy, a bump is observed in the $\gamma$-strength function of rare earth nuclei from ($^3$He,$\alpha$) experiments~\cite{voinov}, and is probably of the same origin as the pygmy resonance found in $(n,\gamma)$ reactions~\cite{iga}.

Section~II describes the experimental methods.
In Sect.~III the experimental level density and $\gamma$-strength function of  $^{166}$Er and $^{167}$Er are obtained.
Section~IV examines thermodynamic properties within the micro-canonical and the canonical ensemble, while electromagnetic properties of the two nuclei are discussed in Sect.~V.
Concluding remarks are given in Sect.~VI.

\section{Experimental method and techniques}
The experiment was carried out at the Oslo Cyclotron Laboratory.
The reactions employed were the neutron pick-up ($^3$He,$\alpha \gamma$) and the inelastic scattering ($^3$He,$^3$He'$\gamma$) with a beam energy of $45$ MeV, populating $^{166}$Er and $^{167}$Er with high excitation energy and low spin.
The $Q$-value of the ($^3$He,$\alpha$)-reaction is $14.142$ MeV.
The self-supporting $^{167}$Er target was isotopically enriched to $95.6$\% and had a thickness of $1.5$ mg/cm$^2$.

The charged particles and $\gamma$ rays were recorded with the detector array
CACTUS~\cite{mg90}, which contains 
8 Si particle telescopes and 27 NaI $\gamma$-ray detectors.
Each telescope is placed at an angle of 45$^{\circ}$ relative to the beam axis, and comprises one Si front and one Si(Li) back detector with thickness $140$ and $3000$ {\normalfont $\mu$}m, respectively.
The NaI $\gamma$-detector array, having a resolution of $\sim 6$ \% at $\gamma$ energy $E_{\gamma} = 1$ MeV and a total efficiency of $\sim15$ \%, surrounds the target and particle detectors.
In addition, two Ge detectors were used to monitor the spin distribution and selectivity of the reactions.

\begin{figure}
\centering
\includegraphics[width=15cm]{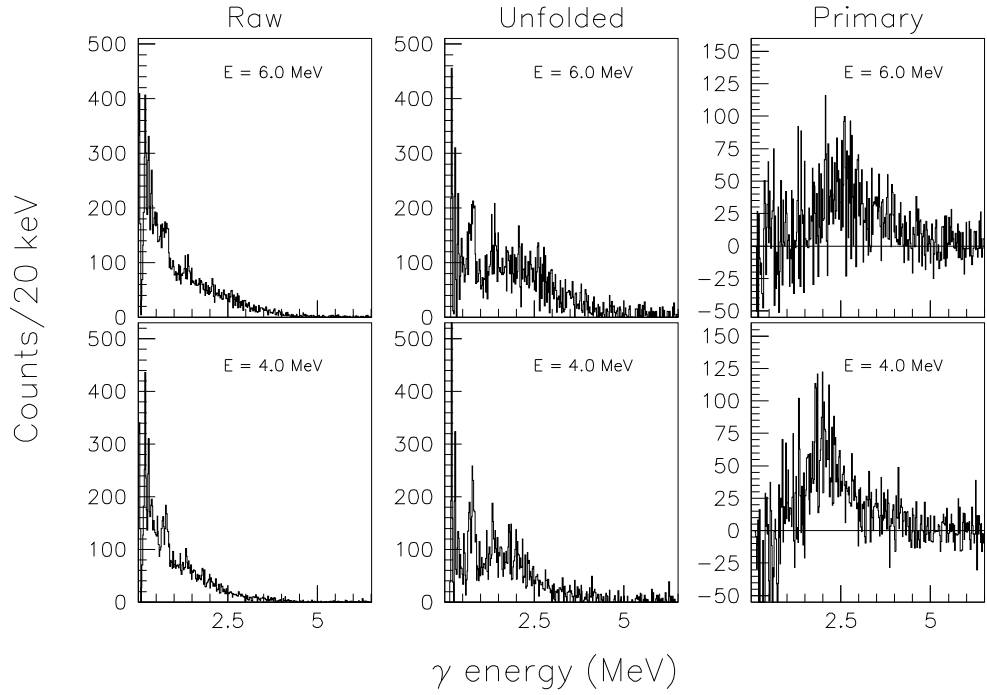}
\caption{Gamma-ray spectra at $E=4$ and $6$ MeV of excitation energy for $^{166}$Er. 
Left panel: Raw $\gamma$-ray spectra, central panel: Unfolded $\gamma$-ray spectra, and right panel: Primary $\gamma$-ray spectra.}
\label{fig:adata}
\end{figure}

\begin{figure}
\centering
\includegraphics[width=17cm]{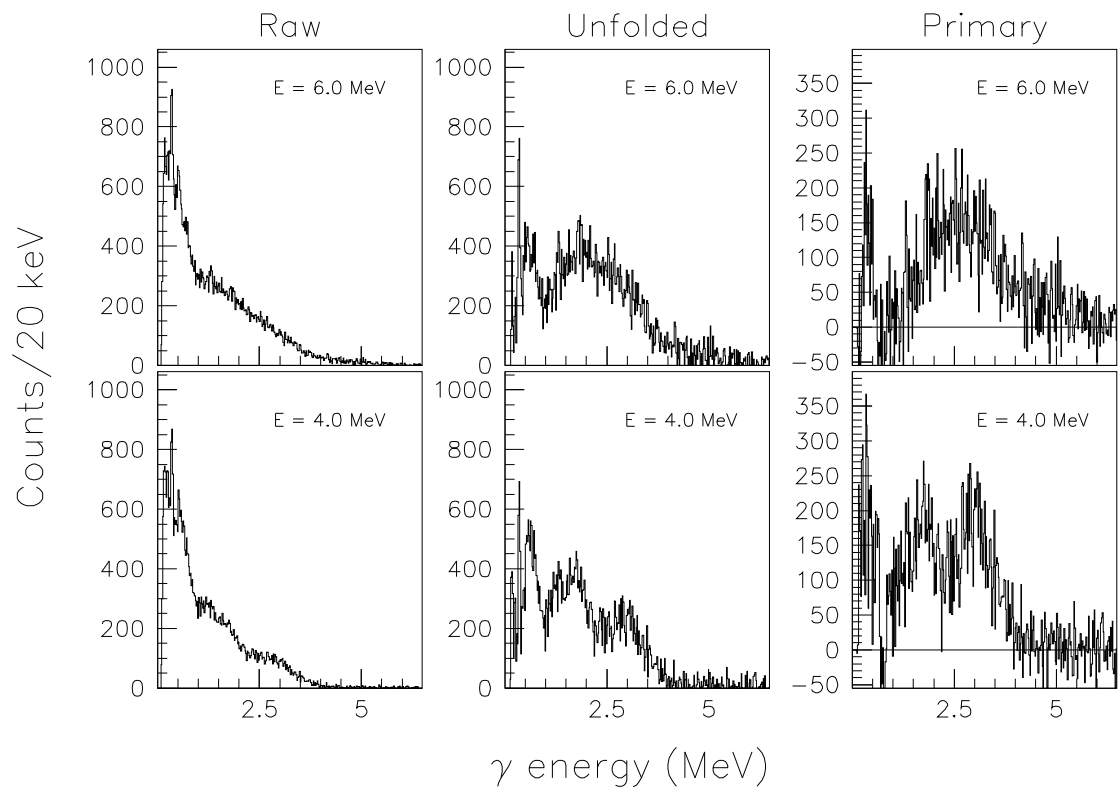}
\caption{Same as Fig.~1 for $^{167}$Er.}
\label{fig:tdata}
\end{figure}

The excitation energy of the resulting $^{166}$Er and $^{167}$Er nuclei are determined by means of reaction kinematics of the ejectile.
By setting proper gates in the particle spectra, each coincident
$\gamma$ ray can be assigned to a $\gamma$ cascade originating from a specific excitation energy. The data are sorted into a matrix of $(E,E_{\gamma})$ energy pairs, $E$ being the excitation energy.
Examples of the recorded $\gamma$ spectra, the so called raw
spectra, from two excitation energies are shown in the left panel of Figs.~\ref{fig:adata} and~\ref{fig:tdata} for $^{166}$Er and $^{167}$Er, respectively.
Note that the statistics of $^{167}$Er is about twice as good as for $^{166}$Er, since the ($^3$He,$^3$He') reaction has a higher cross section than the ($^3$He,$\alpha$) reaction.
 
In order to determine the true $\gamma$-energy distribution, the $\gamma$ spectra are corrected for the response of the NaI detectors with the unfolding procedure of Ref.~\cite{mg96}. 
Unfolded $\gamma$ spectra are shown in the central panel of Figs.~\ref{fig:adata} and~\ref{fig:tdata} for the two nuclei.

The now corrected $(E,E_{\gamma})$ matrix comprises the $\gamma$-energy distribution of the total $\gamma$ cascade, and make it possible to derive the primary $\gamma$ matrix according to the subtraction technique of Refs.~\cite{mg87,lisa}.
Primary $\gamma$ spectra can be seen in the right panel of Figs.~\ref{fig:adata} and~\ref{fig:tdata}. 

The method of extracting the primary $\gamma$ spectra is based on the assumption that the decay properties of a bin of excited states are independent of whether the states are directly populated through the nuclear reaction or from de-excitation from higher excited states.
This is believed to be approximately fulfilled because the long life time of excited states give the nucleus time to thermalize prior to the $\gamma$ decay.

\section{Extraction of level density and $\gamma$-ray strength function}
The primary $\gamma$ matrix provides information on both the level density and the $\gamma$-ray strength function, enabling a simultaneous determination of the two functions.
The fundamental assumption behind the extraction procedure is the Brink-Axel hypothesis~\cite{brink,axel}, where the probability of $\gamma$-decay in the
statistical regime, represented by the primary $\gamma$ matrix $P(E_i, E_{\gamma})$, can be expressed simply as a product of the
final-state level density $\rho(E-E_{\gamma})$ and a $\gamma$-energy dependent factor
$F(E_{\gamma})$
\begin{equation}
P(E, E_{\gamma}) \propto F(E_{\gamma}) \rho(E-E_{\gamma}).
\label{eq:ab}
\end{equation}
The $\gamma$-energy dependent factor is proportional to $\sum_{XL}E_{\gamma}^{2L + 1}f_{XL}(E_{\gamma})$, where $f_{XL}(E_{\gamma})$ is the $\gamma$-ray strength function for the multipolarity $XL$.
 
During the last few years a method for simultaneous deduction of level density and $\gamma$-strength function from primary $\gamma$ spectra has been developed at the Oslo Cyclotron Laboratory~\cite{andreas:nim,lisa,trine}.
In the previously published results on $^{166}$Er~\cite{elin}, the level densities and $\gamma$-strength functions were still unnormalized. 
The procedure is now fully replaced by an iteration technique deriving the level density and the $\gamma$-energy dependent factor by a $\chi^2$-fit to the experimental primary $\gamma$ spectra~\cite{andreas:nim}.
This technique is described in detail in Ref.~\cite{andreas:nim}, and only an outline of the normalization procedures is given here.
It is important to note that the extracted experimental quantity $\rho$ is in fact the density of levels accessible to the nuclear system in the $\gamma$-decay process.
This quantity is interpreted as the level density, but may be influenced by selection rules in the $\gamma$ decay.
For a thermalized nucleus in the continuum, the interpretation is approximately valid~\cite{andreas61}.

Equation~(\ref{eq:ab}) has an infinite number of solutions.
It can be shown~\cite{andreas:nim} that all equally good solutions of Eq.~(\ref{eq:ab}) can be obtained by the transformations
\begin{eqnarray}
\tilde{\rho}(E-E_{\gamma}) &=& A \exp[\alpha(E-E_{\gamma})] \rho(E-E_{\gamma}) , 
\label{eq:array}
\\
\tilde{F}(E_{\gamma}) &=& B \exp(\alpha E_{\gamma}) F(E_{\gamma}) 
\label{eq:array2}
\end{eqnarray} 
of any particular solution ($\rho,F$).
Consequently, neither the slope nor the absolute value of the two functions can be obtained through the iteration procedure, but the three variables $A$, $B$ and $\alpha$ of Eqs.~(\ref{eq:array}) and~(\ref{eq:array2}) have to be determined independently to give the best physical solution of the level density and $\gamma$-strength function.

\subsection{The level density}
\begin{figure}
\centering
\includegraphics[width=17cm]{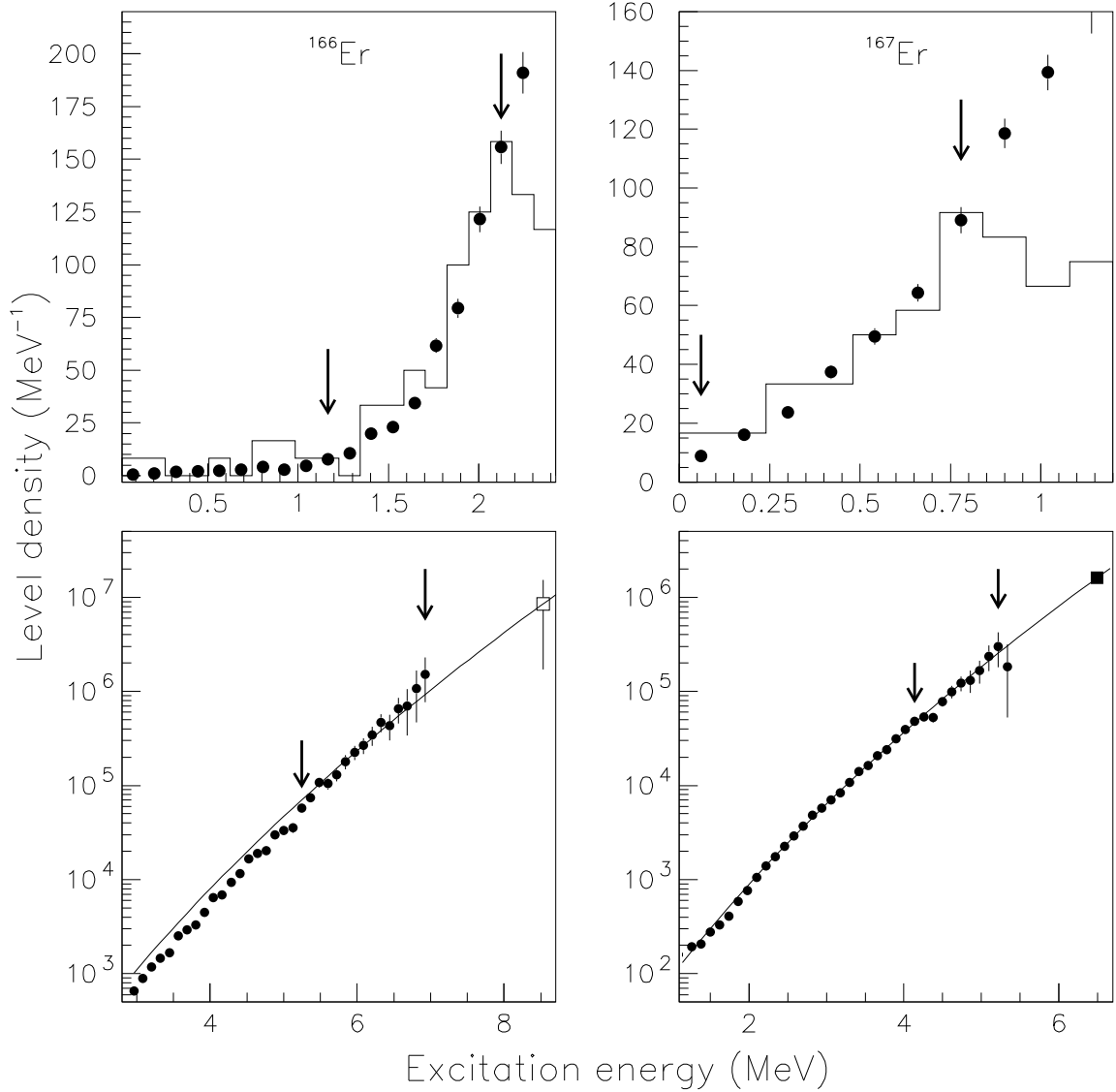}
\caption{Normalization of the experimental level density (data points) of $^{166}$Er (left panel) and $^{167}$Er (right panel) between the arrows to known levels at low excitation energy (histograms) and to the Fermi-gas level density ($^{166}$Er) or to the level density calculated from neutron-resonance spacings ($^{167}$Er) at the neutron-separation energy (squares).
The lines are the Fermi-gas approximations.}
\label{fig:normrho}
\end{figure}

The parameters $A$ and $\alpha$ of Eq.~(\ref{eq:array}) can be determined by fitting the level density from the iteration procedure to the number of known discrete levels~\cite{toi} at low excitation energy and to the level density estimated from neutron-resonance spacing data at high excitation energy~\cite{ilj92}.
This normalization procedure is shown for $^{167}$Er in the right panel of Fig.~\ref{fig:normrho}.
The deduced level density is fitted to the discrete levels (between the arrows in the upper panel of Fig.~\ref{fig:normrho}) as far up in energy as we can assume that all levels are known.
At high excitation energies the deduced level density is fitted (between the arrows in the lower panel of Fig.~\ref{fig:normrho}) to a Fermi-gas approximation of the level density (line) forced to pass through the level-density estimate at the neutron-separation energy (filled square) obtained from the neutron-resonance spacing data.

Unfortunately, $^{165}$Er is an unstable nucleus ($T_{1/2}=10.36$ h), making the neutron capture $^{165}$Er(n,$\gamma$)$^{166}$Er difficult to investigate. 
Thus, there is no neutron-resonance data available for $^{166}$Er.
In this case the level density at the neutron-separation energy is estimated with the Fermi-gas expression 
\begin{equation} \rho= \frac{\exp(2 \sqrt{aU})}{12 \sqrt{2}a^{1/4}U^{5/4} \sigma}, \end{equation}
where the level density parameter $a=0.21A^{0.87}$, the shifted excitation energy $U=E-C_1-\Delta$, and the spin-cutoff parameter $\sigma$ is defined by $\sigma^2=0.0888\sqrt{aU}A^{2/3}$, while $A$ is the mass number, $\Delta$ the pairing-gap parameter and $C_1=-6.6A^{-0.32}$. 
All parameters are calculated according to Ref.~\cite{egidy}, which provides a level density in acceptable agreement with the experimental level density at the neutron-separation energy for $^{167}$Er. 
Also the nuclei $^{161,162}$Dy and $^{171,172}$Yb show conformity between the experimental and theoretical level density at the neutron-separation energy~\cite{magne62}.
The left panel of Fig.~\ref{fig:normrho} shows the normalization of $^{166}$Er, where the open square represents the theoretically calculated level density at the neutron-separation energy with an error of $80$\%~\cite{egidy}.  

\begin{figure}
\centering
\includegraphics[width=17cm]{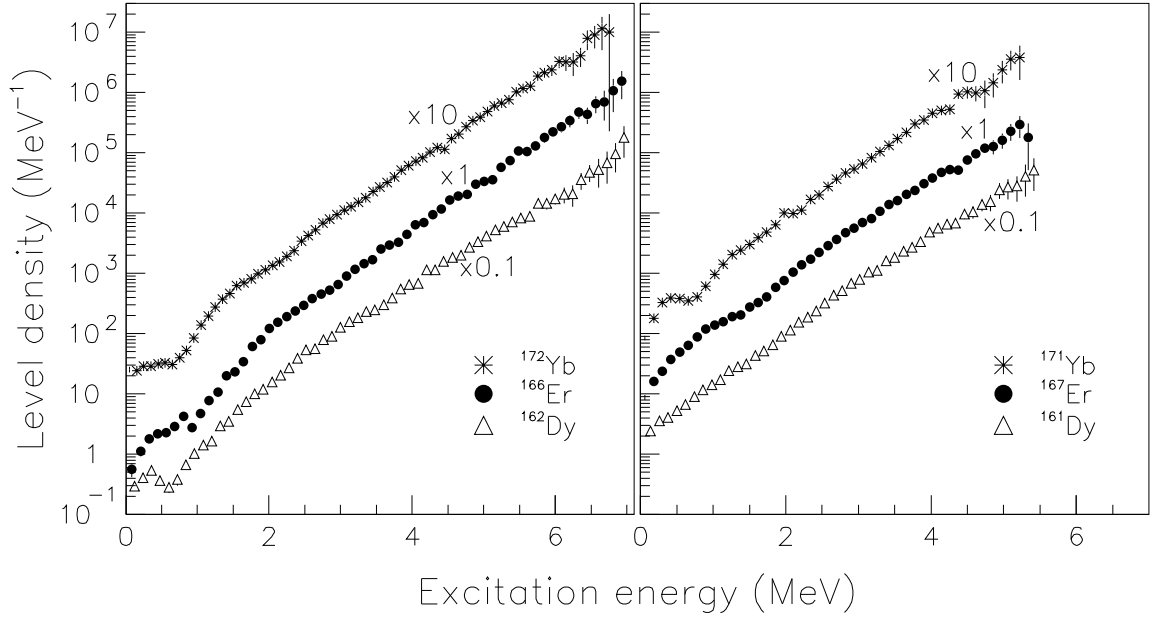}
\caption{
The normalized level density of $^{166}$Er (left) and $^{167}$Er (right) compared to $^{162}$Dy and $^{172}$Yb, and $^{161}$Dy and $^{171}$Yb, respectively.
The Dy isotopes are multiplied by 0.1 and the Yb isotopes by 10 for better visualization.
All nuclei are popolated through the ($^3$He,$\alpha$) reaction except for $^{167}$Er, which is populated through the ($^3$He,$^3$He') reaction.
}
\label{fig:rhosmnl}
\end{figure}

The normalized level density of $^{166}$Er is compared to the level densities of $^{162}$Dy and $^{172}$Yb~\cite{voinov} in the left panel of Fig.~\ref{fig:rhosmnl}.
The three level-density functions coincide well at high excitation energies, indicating that the chosen value for the level density at the neutron-separation energy for $^{166}$Er is reasonable.
For excitation energies below $1.9$ MeV the experimental level density of the erbium nucleus is seen to be lower than for the two others.
In this energy region most levels in the three nuclei are believed to be known.
From counting of known levels~\cite{toi}, it is found that between $1.1$ and $1.9$ MeV, the level density of $^{162}$Dy and $^{172}$Yb indeed exceed the level density of $^{166}$Er with the same degree as observed experimentally.

The right panel of Fig.~\ref{fig:rhosmnl} compares the level density of $^{167}$Er with the level densities of $^{161}$Dy and $^{171}$Yb~\cite{voinov}.
The two latter nuclei were populated through ($^3$He,$\alpha$) reactions.
The $^{171}$Yb-level density is seen to be lower than the two others below $1.1$ MeV.
According to~\cite{toi}, this agrees with reality below $900$ keV.
Above this limit the level densities from counting of known levels start to decrease, making it evident that not all levels are known.
 
In Ref.~\cite{andreas61} the level density of $^{162}$Dy extracted from two different reaction mechanisms are compared.
The primary $\gamma$ spectra from the ($^3$He,$\alpha$) pick-up reaction and the inelastic ($^3$He,$^3$He') reaction are found to provide level densities with very good agreement except below $\sim 1$ MeV of excitation energy.
There, the level density obtained from the inelastic $^3$He scattering overestimates the number of levels by a factor of $\sim 3$, because this reaction populates more collective excitations with similar structure as the ground-state band.
This causes a relatively large $\gamma$-transition rate from the direct populated states to the ground band, which in turn gives an overestimation of the level density there.
The same is observed in the level density of $^{172}$Yb deduced from the inelastic reaction~\cite{andreasdr}.
This effect of overestimation can not be ruled out in the level density of $^{167}$Er derived from the ($^3$He,$^3$He') reaction, however, the experimental data fit the known level density at low excitation energy very well, see Fig.~\ref{fig:normrho}. 
A difference between these ($^3$He,$^3$He') experiments is the spin of the target nuclei.
The $^{167}$Er target has a spin of $7/2$ making excitations to a large number of states probable, with a resulting fragmentation of collective strength and a lower probability for subsequent $\gamma$ transitions directly into the ground-state band than in the even-even nuclei.
Thus, the accessible states in the $\gamma$ decay seem to approximates the level density quite well.

\subsection{The $\gamma$-ray strength function}
Blatt and Weisskopf~\cite{blatt} suggested the ratio of the partial radiative width $\Gamma_i(E_{\gamma})$ and the level spacing of the initial states $i$ with equal spin and parity $D_i$ to describe the $\gamma$ decay in the continuum. 
The corresponding definition of the $\gamma$-ray strength function is given by
$f_{XL}=\Gamma_i(E_{\gamma})/(E_{\gamma} ^{2L+1}D_i)$, where $X$ denotes the electric or magnetic character, and $L$ defines the multipolarity of the $\gamma$ transition. 

\begin{figure}
\centering
\includegraphics[width=17cm]{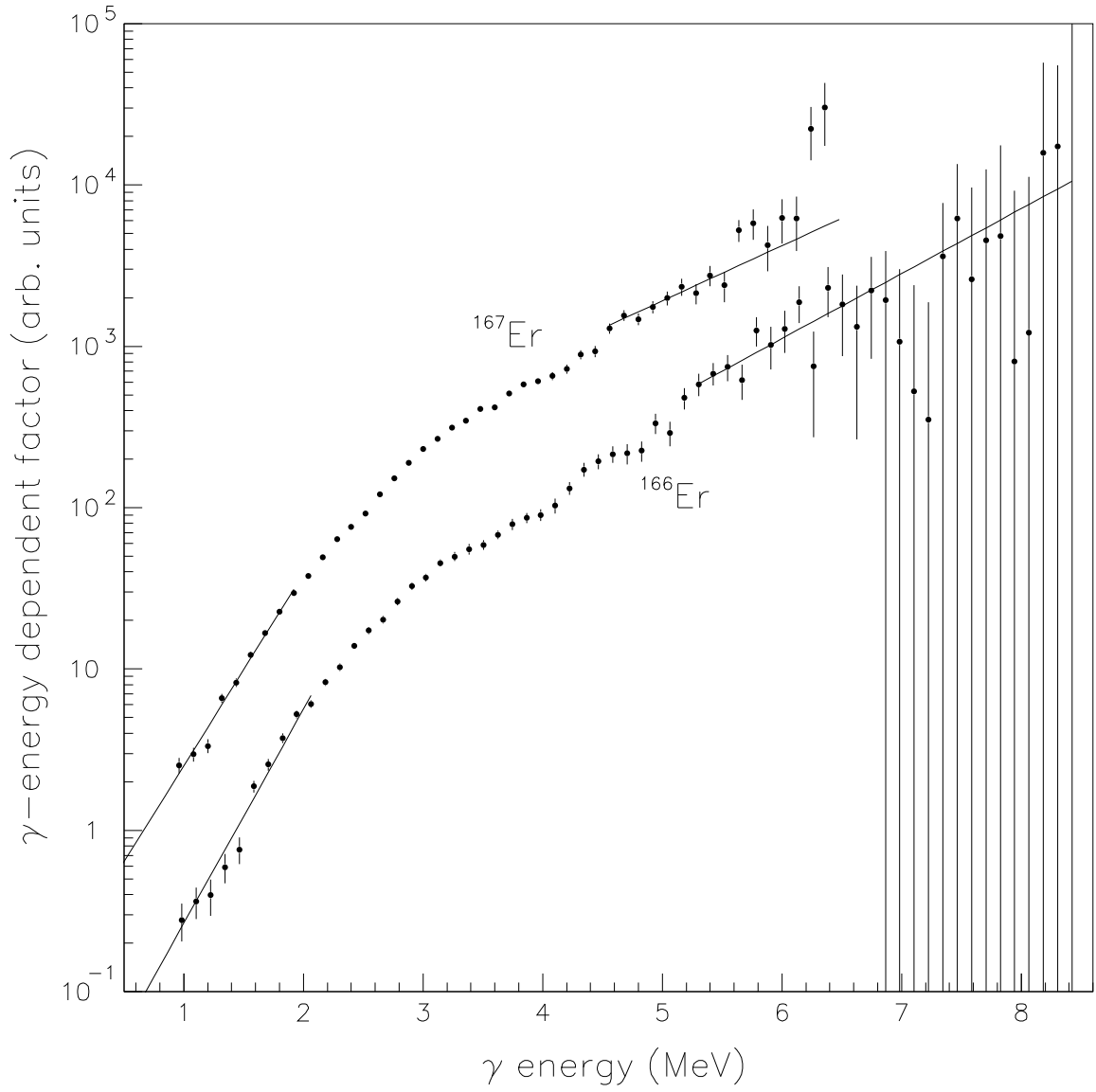}
\caption{The $\gamma$-energy dependent factor of $^{166,167}$Er.
The lines are extrapolations needed to calculate the normalization integral of Eq.~(\ref{eq:norm}).}
\label{fig:f}
\end{figure}

After the normalization of the level density, the parameter $\alpha$ of Eq.~(\ref{eq:array2}) is already fixed, and the slope, i.e.\ the exponential $\exp(\alpha E_{\gamma})$, is included in the $\gamma$-energy dependent functions shown for $^{166}$Er and $^{167}$Er in Fig.~\ref{fig:f}.
Still, the normalization constant $B$ of Eq.~(\ref{eq:array2}) remains to be determined.

The $\gamma$-energy dependent factor $F(E_{\gamma})$ is proportional to the sum of $E_{\gamma}^{2L+1} f_{XL}(E_{\gamma})$ for all possibilities of $X$ and $L$.
We assume that the $\gamma$ decay in the continuum of nuclei with low spin is mainly governed by electric and magnetic dipole radiation and that the accessible levels have equal numbers of positive and negative parity states. 
Thus, the observed $F$ can be expressed by a sum of the E1 and M1 $\gamma$-strength functions only:
\begin{equation}
BF(E_{\gamma})=
[f_{\mathrm{E1}}(E_{\gamma})+f_{\mathrm{M1}}(E_{\gamma})] E_{\gamma}^3.
\label{eq:prim1}
\end{equation}
The  average total radiative width of neutron resonances $\langle\Gamma_\gamma\rangle$~\cite{kop} with excitation energy equal to the neutron-separation energy $S_n$, spin $I$, and parity $\pi$
\begin{eqnarray}
\langle\Gamma_\gamma\rangle=\frac{1}{\rho(S_n, I, \pi)} \sum_{XL}\sum_{I_f, \pi_f}&&\int_0^{S_n}{\mathrm{d}}E_{\gamma} E_{\gamma}^{2L+1} f_{XL}(E_{\gamma})
\nonumber\\
&&\rho(S_n-E_{\gamma}, I_f, \pi_f), 
\label{eq:norm}
\end{eqnarray}
can be written in terms of $F$ by means of Eq.~(\ref{eq:prim1}). 
With the experimental level density $\rho$ already normalized, the normalization constant $B$ can be deduced.
Assuming $s$-neutron capture, $I,\pi$ take the values $I_t\pm \frac{1}{2},\pi_t$, where $I_t,\pi_t$ are the spin and parity of the target nucleus in the $(n,\gamma)$ reaction.
The experimental value of the total radiative width is the average over the possible spins of the compound state ($I_t\pm \frac{1}{2}$). 
Since we expect the $\gamma$ decay to be governed by dipole transitions, the second sum is restricted to possible final states with spin $I_f$ and parity $\pi_f$ accessible by dipole radiation. 

The methodical difficulties in the primary $\gamma$-ray extraction, prevents determination of the functions $F(E_{\gamma})$ and $\rho(E)$ in the interval $E_\gamma<1$~MeV and $E > S_n-1$~MeV, respectively. 
In addition, the data at the highest $\gamma$-energies, above $E_{\gamma} \sim S_n-1$~MeV, suffer from poor statistics. 
Therefore, extrapolations of $F$ and $\rho$ were necessary in order to calculate the integral of Eq.~(\ref{eq:norm}). 
For the level density a Fermi-gas extrapolation is used (see Fig.~\ref{fig:normrho}), and for the $\gamma$-energy dependent factor a pure exponential of the form $\exp(bE_{\gamma})$ is utilized (see Fig.~\ref{fig:f}).
The contribution of the extrapolation to the total radiative width in Eq.~(\ref{eq:norm}) does not exceed $15$\%, thus the errors due to a possibly poor extrapolation are expected to be of minor importance~\cite{voinov}. 

\begin{figure}
\centering
\includegraphics[width=17cm]{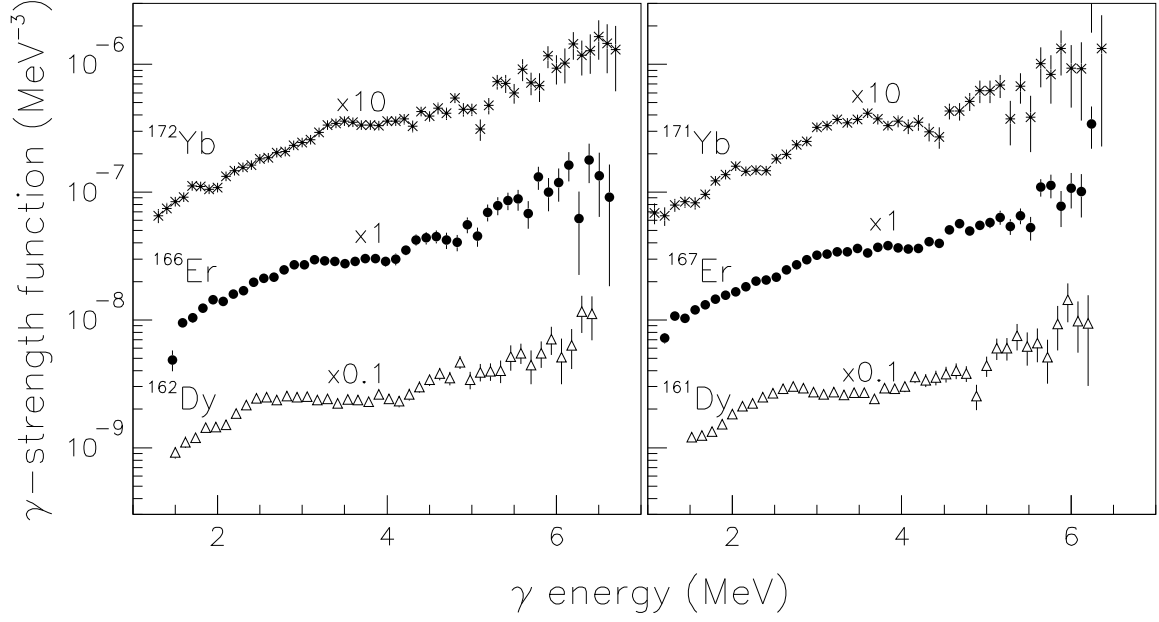}
\caption{The normalized $\gamma$-ray strength function of $^{166}$Er (left) compared to $^{162}$Dy and $^{172}$Yb, and $^{167}$Er (right) compared to $^{161}$Dy and $^{171}$Yb.
The Dy isotopes are multiplied by 0.1 and the Yb isotopes by 10 for better visualization.
}
\label{fig:sfsmnl}
\end{figure}

Values for the neutron-resonance radiative width are given in~\cite{tecdoc}, and for $^{167}$Er we find $\langle\Gamma_\gamma\rangle= 92(8)$ meV. 
Since the radiative width of $^{166}$Er is unknown, its value is taken as the average of $\langle\Gamma_\gamma\rangle$ of the neighbouring isotopes.
This is justified by the general slow variance in the $\langle\Gamma_\gamma\rangle$ values of neighbouring isotopes for other elements.
From interpolation, we thus adopt the value $\langle\Gamma_\gamma\rangle= 90(20)$ meV for $^{166}$Er.
The normalized $\gamma$-strength functions of $^{166}$Er and $^{167}$Er are compared to the strength functions of $^{162}$Dy and $^{172}$Yb, and $^{161}$Dy and $^{171}$Yb, respectively, in Fig.~\ref{fig:sfsmnl}. 
All strength functions are increasing smoothly with $\gamma$ energy except for a bump around $E_{\gamma}\sim3$ MeV.
The $\gamma$-strength functions of neighbouring isotopes display a striking qualitative similarity.
This can be expected for nuclei with approximately the same charge distribution. 
The location of the bump can however be seen to move towards higher $\gamma$ energies with increasing mass number.   

\section{Thermodynamic properties}
For a statistical description of hot nuclei, the micro-canonical thermodynamics is the proper theory.
Within this frame, the system is isolated, giving a well defined energy.
However, the canonical ensemble, permitting heat exchange, and the grand-canonical ensemble, which in addition allows particle exchange, are often used due to mathematical difficulties with detailed calculations within the micro-canonical ensemble.
In this work, both the micro-canonical and the canonical ensemble will be utilized to discuss thermodynamic properties of the $^{166,167}$Er nuclei experimentally.

In particular we will focus on different ways, within the two ensembles, to obtain experimental values of the critical temperature for the pair-breaking process and the general quenching of pair correlations.
The micro-canonical ensemble gives a detailed description of the breaking of one, two, three,... nucleon pairs as a function of excitation energy, while the canonical ensemble reveals the general average properties of this phase-like transition.
The proper definition of thermally driven first- and second-order phase transitions in systems with few particles is a long standing problem, which will not be discussed in the present experimental work.

\subsection{Micro-canonical ensemble}
The micro-canonical partition function is simply the multiplicity of nuclear states, which experimentally corresponds to the level density of accessible states.
Thus, the experimental level density $\rho(E)$ is our starting point for the extraction of thermodynamic properties of nuclei.

The entropy is determined by
\begin{equation}
S(E)=S_0 + \ln \rho(E),
\end{equation}
where the Boltzmann constant for simplicity is set to unity ($k_B\equiv 1$), and the normalization constant $S_0$ can be adjusted to fulfill the condition of the third law of thermodynamics; $S\rightarrow 0$ when $T\rightarrow 0$, $T$ being the nuclear temperature
\begin{equation}
T(E)=\left(\frac{\partial S}{\partial E}\right)^{-1}.
\label{eq:t}
\end{equation}
The ground-state band of even-even nuclei is assumed to have $T=0$.
Therefore, $S_0$ is determined so that the entropy of the ground-state band in $^{166}$Er is approximately zero.

\begin{figure}
\centering
\includegraphics[width=17cm]{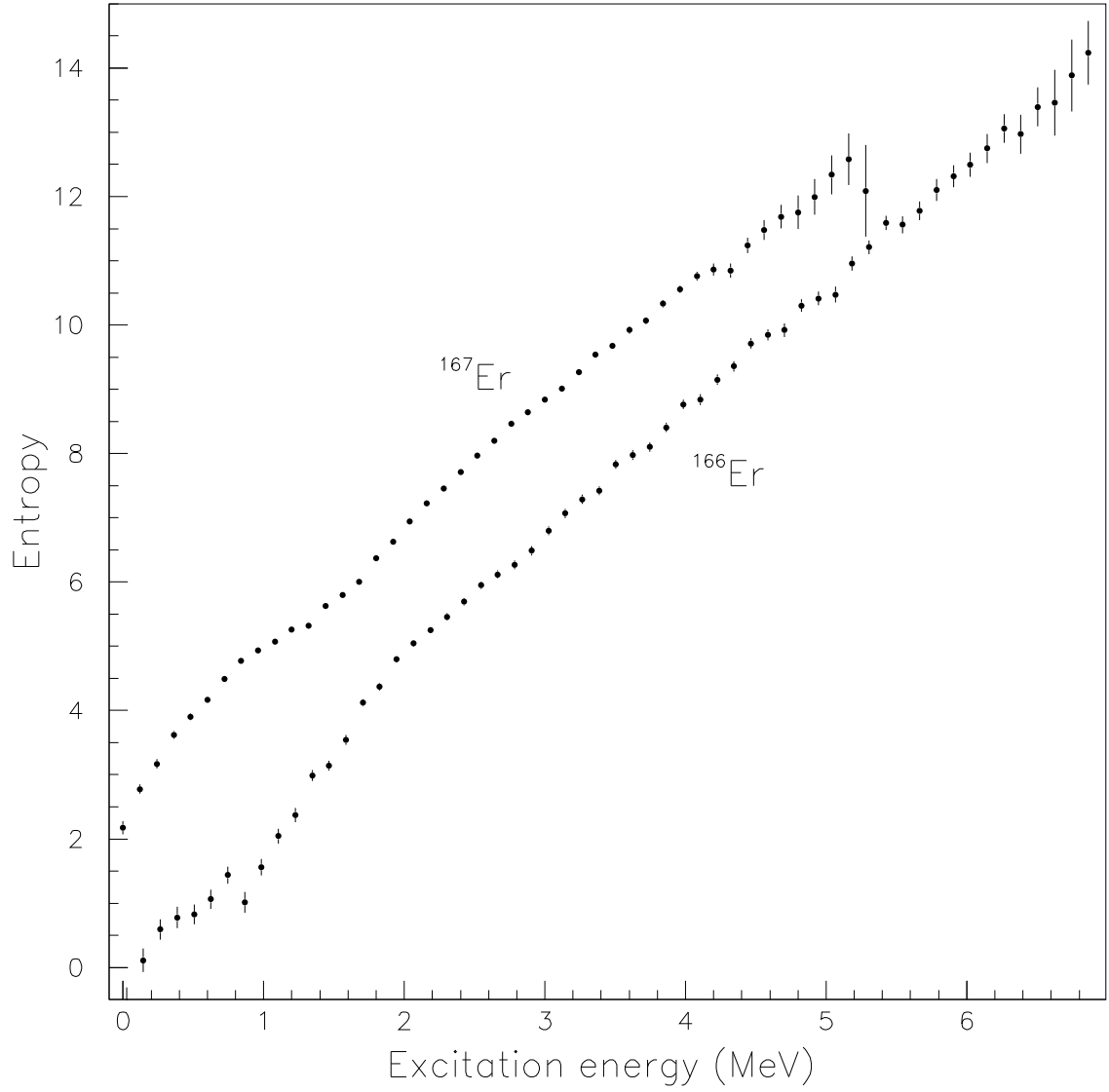}
\caption{The entropy as a function of excitation energy in $^{166}$Er and $^{167}$Er}
\label{fig:s}
\end{figure}

\begin{figure}
\centering
\includegraphics[width=17cm]{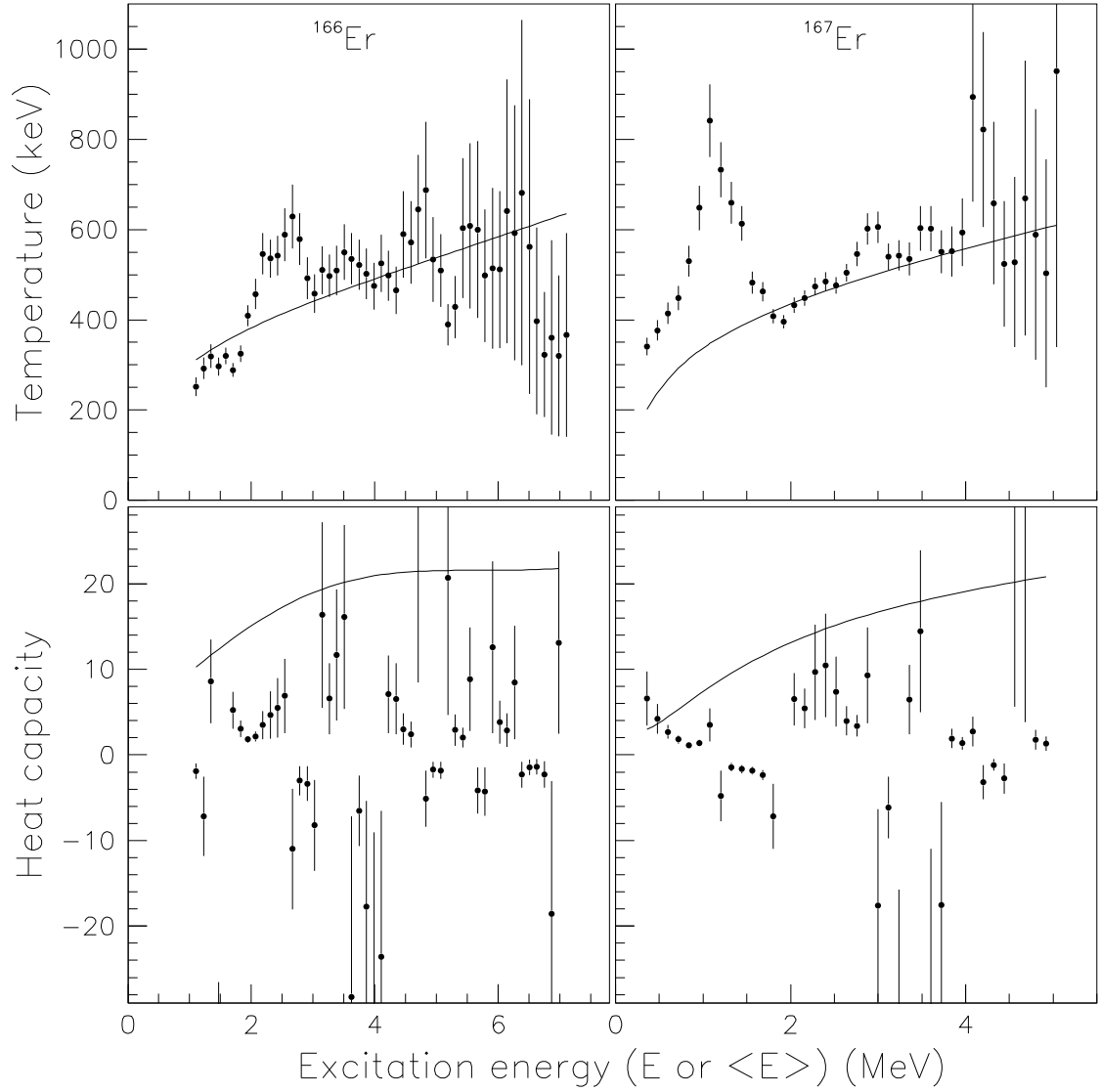}
\caption{The temperature (upper panel) and heat capacity (lower panel) of $^{166}$Er (left) and $^{167}$Er (right) from the micro-canonical
ensemble (data points) and the canonical ensemble (lines).
}
\label{fig:mu}
\end{figure}

Figures~\ref{fig:s} and~\ref{fig:mu} show the entropy and the temperature, respectively, deduced in the micro-canonical ensemble for $^{166}$Er and $^{167}$Er.
The entropy curve plotted in a linear scale is essentially identical to the level-density curve in a logarithmic scale.
The small bumps in the entropy curves of Fig.~\ref{fig:s} are enhanced through the differentiation performed in Eq.~(\ref{eq:t}) to obtain the temperatures of Fig.~\ref{fig:mu}. 

The specific heat can be determined from differentiating the temperature
\begin{equation}
C_V(E) =\left(\frac{\partial T}{\partial E}\right)^{-1}
\end{equation}
and is displayed in the lower panel of Fig.~\ref{fig:mu} for $^{166}$Er and $^{167}$Er.
The double differentiation of the entropy has introduced strong fluctuations in the specific heat. 
Still it is distinct that the oscillations in the temperature give rise to successive positive and negative heat capacity.
The spectacular feature of negative heat capacity is a direct consequence of the decrease in the micro-canonical temperature, and has recently been observed experimentally in the critical region of nuclear fragmentation in Au quasi-projectile sources formed in Au+Au collisions~\cite{agost}.

An anomalous decrease in the temperature means an unusual increase in the entropy, i.e.~a more than normal opening of new domains of the phase space, e.g.~opening of new degrees of freedom as more particle pairs are broken.
The successive bumps of Figs.~\ref{fig:s} and~\ref{fig:mu} indicate that the transition from a paired to an unpaired phase is a gradual process of breaking up more and more nucleon pairs.

When we interpret the area of negative heat capacity, corresponding to the region where the temperature has a negative slope, as the location of the break up of a nucleon pair, a signal of the break up of the first pair can be seen around $E\sim 1.5$ MeV in $^{167}$Er in Fig.~\ref{fig:mu}.
This value is close to twice the pairing-gap parameters $2\Delta_n=1.7$ MeV and $2\Delta_p=1.5$ MeV, which are the expected cost of breaking of a neutron and proton pair, respectively.
The pairing-gap parameters are calculated from the empirical masses of a sequence of isotopes or isotones~\cite{bm1} of $^{167}$Er.
The first negative slope in the temperature as a function of excitation energy in $^{161}$Dy, $^{162}$Dy, $^{171}$Yb, and $^{172}$Yb~\cite{andreasdubna} also coincide roughly with $2\Delta$.  
A similar argument is however less successful for $^{166}$Er.
Deviations from $2\Delta$ in the localization of the break up of the first pair of particles can be due to the influence of several structural effects in the nuclei, such as e.g.~the Fermi-level position in the Nilsson single-particle scheme, variation in the density of single-particle orbitals, and two quasi-particle couplings to collective degrees of freedom.

It has been shown~\cite{gross3} that phase transitions of finite-size systems can be observed in the micro-canonical ensemble without invoking the thermodynamic limit.
According to~\cite{gross}, a phase transition of first order can be recognized from the caloric curve in the region of co-existing phases, i.e.~in the excitation energy region where the phase transition takes place.
There, both the phase the system is leaving and the phase the system is attending are existing simultaneously.
For this co-existency the system has to ``pay'' an amount of entropy which is eventually returned as the whole system is converted into the new phase~\cite{gross}.
The non-oriented area between the inverse temperature $T(E)^{-1}$ and the line $T_c^{-1}$ is twice this entropy.
The critical temperature has to satisfy the condition that the oriented area between $T(E)^{-1}$ and $T_c^{-1}$ equals zero.

The process studied in the present work, namely the breaking of nucleon pairs and the general quenching of pair correlations, deviates from the process of multifragmentation~\cite{gross3,gross}.
In the latter process, the geometrical surfaces of the fragments play a crucial role.
In the depairing process, the long-range two-nucleon force has correlation lengths longer than the diameter of the nucleus.
Thus, the phase-like situation of having a pair coupled or not, reveals no geometrical surface of interaction.

\begin{figure}
\centering
\includegraphics[width=10cm]{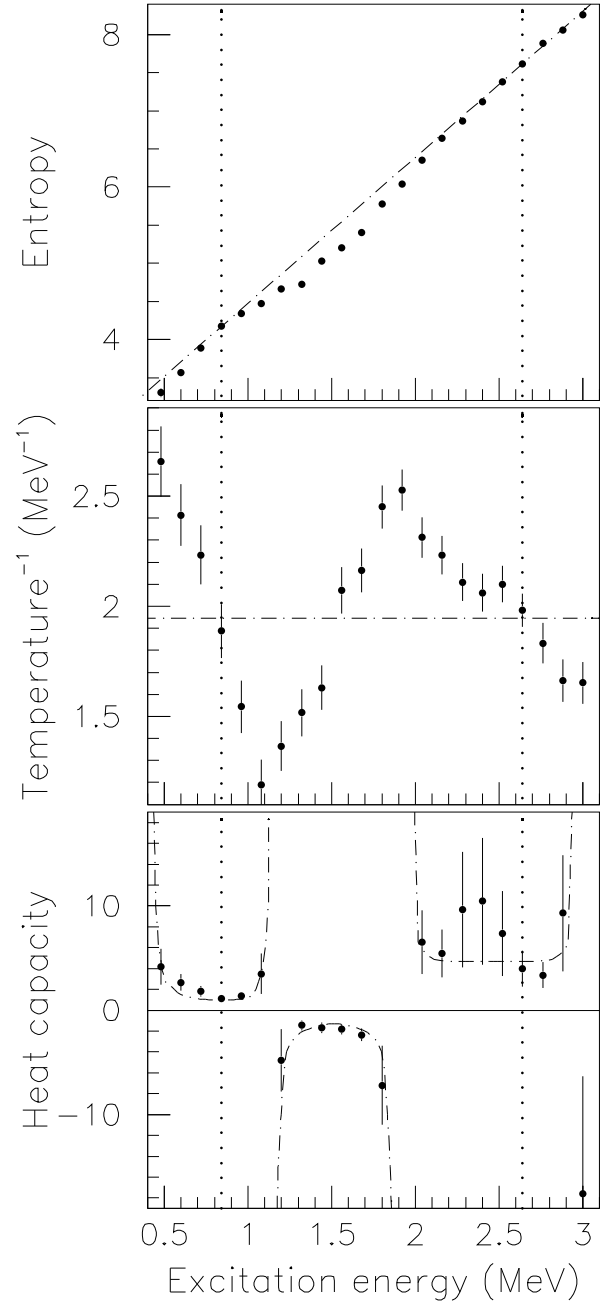}
\mbox{}
\caption{Entropy (upper panel), inverse temperature (central panel) and specific heat (lower panel) as functions of excitation energy in $^{167}$Er.
The vertical, dotted lines indicate the approximate region in which the phase-like transition takes place.
See text.}
\label{fig:gross}
\end{figure}

In order to qualitatively analyze our experimental findings along the procedure of Ref.~\cite{gross}, the excitation energy region around the outstanding bump in the temperature of $^{167}$Er is expanded, and the entropy, inverse temperature and heat capacity of $^{167}$Er are plotted in Fig.~\ref{fig:gross}.
The qualitative similarity with the picture of Ref.~\cite{gross} suggests that the general process of breaking a nucleon pair around $E\sim 2\Delta$ appears like a phase transition of first order.

Accepting this theoretical result, however, the similar experimental signals of breaking of further nucleon pairs indicate that the process of quenching of pairing correlations is a series of first order phase transitions.
This interpretation might seem physically unattractive, and therefore we are anticipating further theoretical clarifications.

The convex shape of the entropy curve of Fig.~\ref{fig:gross} indicates a transition from the one quasi-particle regime to the three quasi-particle regime.
The transition takes place between the two tangent points of the entropy and its enveloping curve (dashed line in the entropy plot).
The transition region is marked with two dotted vertical lines.
The critical temperature for breaking of the first nucleon pair in $^{167}$Er is derived to be $T_c=0.51(4)$ MeV (horizontal dashed line in the caloric curve of Fig.~\ref{fig:gross}).

The inter-phase entropy\footnote{As stated above, these phases are not physically separated.} could in principle be expected to be equal to the envelope entropy, that is, an interpolation of the entropies before and after the pair-breaking process.
But compared to this expectation, the entropy is seen to be lowered by $\sim 0.6$ units, corresponding to the sum of the primary loss and the final gain in entropy through the transition.
It is unclear how this entropy-mixing contribution shall be interpreted in a small nuclear system.

The first law of thermodynamics states that supplied heat equals the change in internal energy plus the work performed by the system: $dQ=dE+dW$.
Since the nucleus is an isolated system the work $dW$ is zero, and the increase in energy equals the supplied heat through the phase transition.
In macroscopic systems, where the temperature is constant through a phase transition of first order, this increase in energy is the latent heat.
We interpret the transition-region energy  of approximately $1.8$ MeV (see Fig.~\ref{fig:gross}) as the latent heat, even though the temperature is not constant.
We note that this value is also very close to the expected cost $2\Delta$, of breaking up a particle pair.
The dashed lines in the specific heat are purely meant as guiding lines towards the theoretical prediction of poles between the successive positive and negative branches.

The derivative of the entropy-envelope curve in Fig.~\ref{fig:gross} equals the critical temperature.
The somewhat stronger concave shape of the micro-canonical entropy of $^{167}$Er compared to $^{166}$Er (see Fig.~\ref{fig:s}) suggests that increasing temperature is necessary to break further nucleon pairs of $^{167}$Er, whereas the critical temperature for breaking of particle pairs is rather constant in $^{166}$Er.
After the break up of the first particle pair in $^{167}$Er, the slope of the entropy becomes however more constant. 
A linear fit to the entropy above $E\sim 2$ MeV gives a critical temperature of $0.53(10)$ MeV for $^{166}$Er and $0.55(4)$ MeV for $^{167}$Er, indicating that the critical temperature for breaking of nucleon pairs in $^{167}$Er has a slight increase. 
This can be understood from the blocking effect of the unpaired nucleon, increasing the distance to the Fermi surface for low-lying orbitals with coupled pairs. 

The critical temperature for breaking of the first particle pair in $^{167}$Er could be nicely determined from equal areas in the caloric plot.
Unfortunately, the thermodynamic quantities of small systems are strongly fluctuating.
Thus, the critical temperature for each back bend of the temperature, e.i. breaking of a particle pair, can generally be difficult to obtain by this method.
The slope of the entropy-envelope curve can however always serve as a measure. 

\subsection{Canonical ensemble}
The transformation from the micro-canonical to the canonical ensemble is performed by the canonical partition function 
\begin{equation}
Z(T)=\sum_{E=0}^{\infty}\omega (E)e^{-E/T}.
\label{eq:z}
\end{equation}
The partition function is a Laplace transform of the multiplicity of states $\omega (E)=\Delta E\rho (E)$, where $\rho (E)$ is the level density of accessible states in the nuclear reaction at the discrete energy $E$ given in energy bins $\Delta E$.

The thermal average of the excitation energy in the canonical ensemble is
\begin{equation}
\langle E(T)\rangle =Z^{-1}\sum_{E=0}^{\infty}E \omega (E)e^{-E/T}.
\label{eq:e}
\end{equation}
By the Laplace transform in Eq.~(\ref{eq:z}) much of the information contained in the micro-canonical level density becomes smeared out.
This smoothing effect can be quantified by the standard deviation for the thermal average of the energy
\begin{equation}
\sigma_{\scriptscriptstyle E}= \sqrt{\langle E^2\rangle -\langle E\rangle ^2},
\end{equation}
which e.g.~gives $\sigma_E= 3$ MeV at $E=7$ MeV.
Thus, fine structure in the thermodynamic observables in the micro-canonical ensemble will not be visible in the canonical ensemble.
The lines in the upper panel of Fig.~\ref{fig:mu} display the smooth variance of the canonical temperature as a function of the thermal average of the excitation energy.

The mathematical justification of Eqs.~(\ref{eq:z}) and~(\ref{eq:e}) is that the summation is performed to infinity.
However, for the typical temperatures studied here, $T<1$ MeV, the level density should be known up to $40$ MeV, which was found~\cite{andreas} to be a sufficiently high upper limit for the summation.
The experimental level density is only covering the excitation energy region from zero to about $S_n-1$ MeV.
In the region of and above the neutron-separation energy $S_n$, the Fermi-gas model is believed to describe the nuclear properties.
Therefore, the experimental level density is extrapolated to higher excitation energies with the Fermi-gas approximation of Ref.~\cite{egidy}, see the solid lines of Fig.~\ref{fig:normrho}.

\begin{figure}[p]
\centering
\includegraphics[width=17cm]{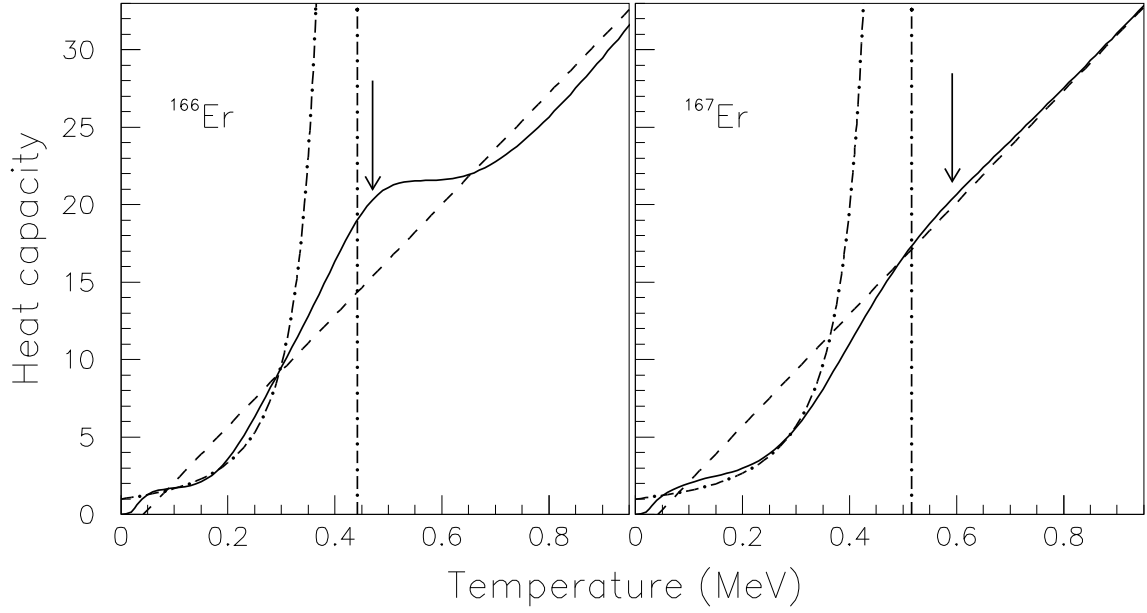}
\caption{The semi-experimental heat capacity of $^{166}$Er (left) and $^{167}$Er (right) as a function of temperature.
The dashed lines are the Fermi-gas approximations $C_V=2aT-3/2$.
The local maxima relative to the Fermi-gas expressions are marked with arrows, while the dashed-dotted lines describe the estimates of Eq.~(\ref{eq:cvt}), where $\tau$ is recognized as the critical temperature and marked with the vertival lines.}
\label{fig:crit}
\end{figure}

The heat capacity in the canonical ensemble is the derivative of the thermal average of the excitation energy
\begin{equation}
C_V(T)=\frac{\partial <E>}{\partial T},
\end{equation}
and it is shown for $^{166,167}$Er as a function of the thermal average of the excitation energy in Fig.~\ref{fig:mu} and as a function of temperature in Fig.~\ref{fig:crit}.
The averaging done by the partition function can be seen to provide a smooth energy and temperature dependence of the heat capacity.
The heat capacity of both nuclei shows an $S$ shape as a function of temperature.
This feature is interpreted~\cite{andreas} as a fingerprint of a phase transition from a state with strong pairing interaction to a state where the pairing correlations are quenched.
Because of the smoothing performed by the canonical-partition function, discrete transitions between the different quasi-particle regimes, as observed within the micro-canonical ensemble, are hidden, and only the phase transition related to the quenching of the pair correlations as a whole can be seen.

The shape of the heat-capacity curve is related to the level-density.
In general a constant-temperature level density gives a pronounced $S$ shape~\cite{andreas}, while a pure Fermi-gas level density provides a linear heat capacity $C_V = 2aT-3/2$; see the dashed lines of Fig.~\ref{fig:crit}.
The more prominent $S$ shape in $^{166}$Er is due to a quicker rise in the heat capacity.
This is the same phenomenon as observed in the micro-canonical ensemble, where the concave shape of the $^{167}$Er entropy was assigned to the blocking effect of the odd neutron.
Shell model Monte Carlo calculations~\cite{alha} on various Fe isotopes have shown that the pairing-phase transition is strongly correlated with the suppression of neutron pairs with increasing temperature.
It is also observed that the reduction of neutron pairs is much stronger in the even- than in the odd-mass isotopes, giving rise to a more pronounced $S$ shape in the even nuclei.
The same difference in the heat capacity is also observed experimentally between $^{161}$Dy and $^{162}$Dy, and $^{171}$Yb and $^{172}$Yb~\cite{andreas}.

In order to extract a critical temperature for the quenching of pair correlations from the canonical data, two different approaches has been tested.
First, the local maximum of the heat capacity relative to the Fermi-gas approximation of Fig.~\ref{fig:crit} have been determined to be $0.47(10)$ MeV for $^{166}$Er and $0.59(10)$ MeV for $^{167}$Er, see the arrows of Fig.~\ref{fig:crit}.
This method relies very much on the Fermi-gas extrapolation of the level density. 
The second approach depends on the assumption that the nuclear level density can be approximated by a constant-temperature expression at low excitation energies.
Within the canonical ensemble this assumption provides the relations 
\begin{equation}T^{-1}=\langle E(T)\rangle^{-1}+\tau^{-1}, 
\label{eq:invT}
\end{equation}
\begin{equation}C_V(T)=(1-T/\tau)^{-2},
\label{eq:cvt}
\end{equation}
where $\tau$ is the constant-temperature parameter.
By fitting $T^{-1}$ as a function of $\langle E(T)\rangle^{-1}$ to a straight line of slope $1$, $\tau$ can be determined.
To minimize the dependency of the extracted critical temperature on the extrapolated level density, the fit is performed between $\langle E(T)\rangle\sim 0.5-2$ MeV.
According to Eq.~(\ref{eq:e}) this corresponds to energies in the level density curve up to $\sim 6$ MeV.
Then $\tau$ is identified with the critical temperature, since $C_V(T)$ has a pole at $T=\tau$. 
The critical temperature is determined to $T_c=0.44(10)$ MeV for $^{166}$Er and $T_c=0.52(9)$ MeV for $^{167}$Er.
The dashed-dotted lines of Fig.~\ref{fig:crit} describe Eq.~(\ref{eq:cvt}) with asymptotes at $\tau=T_c$.

The different values derived for $T_c$ are quoted in Table~\ref{tab:crit}.
The micro-canonical caloric-curve method (Fig.~\ref{fig:gross}) is only supposed to provide the critical temperature for breaking of the first particle pair, while the others should give the critical temperature for the total quenching of the pairing correlations.
The canonical constant-temperature approach relies on experimental data, but it can be questionable whether Eqs.~(\ref{eq:invT}) and~(\ref{eq:cvt}) describes the nucleus well at low excitation energies.
The method of deducing $T_c$ from the local maximum of the canonical heat capacity relative to the Fermi-gas approximation depends highly on the extrapolation of the level density.
The most straight forward method is to obtain the temperature directly from the slope of the micro-canonical entropy.   
There seem to be a tendency for the critical temperature in $^{166}$Er to be slightly lower than in $^{167}$Er.
All results are however equal within the approximated uncertainties, with a mean value of $0.5(1)$ MeV for both nuclei.

The results for $^{166}$Er agrees with recent finite-temperature Hartree Fock Bogoliubov calculations for $^{164}$Er~\cite{egido}, in which the critical-temperature estimate extracted is $0.7$ MeV.
This value is somewhat higher than the present derived result for $^{166}$Er.
This diversity is probably due to application of different interpretations of where exactly the phase transition takes place.
Since the shapes of the heat-capacity curves of $^{164}$Er and $^{166}$Er are practically identical, we assume that the calculated critical temperature for quenching of pair correlations in $^{164}$Er coincides with the experimental findings for $^{166}$Er.

Relativistic Hartree-BCS calculations~\cite{agrawal} also find the proton and neutron pairing gaps to vanish around $0.4-0.5$ MeV of temperature in $^{166}$Er and $^{170}$Er. 
The heat capacity shows two peaks corresponding to these events, characteristic of second order phase transitions from superfluid to normal phase~\cite{agrawal}. 
Thus theoretically, the pair quenching is calculated to be a phase transition of second order.
Provided that negative heat capacity is a sufficient signal, the observed breaking of nucleon pairs in the micro-canonical ensemble appears like first-order phase transitions.
The process of breaking one nucleon pair does however not lead to a total quenching of the pair correlations, and may not be described by the same physics as the pairing-phase transition.
It is not clear how a series of phase transitions should be interpreted physically.

\section{Electromagnetic properties}
The $\gamma$-strength function is governed by different multipolarities of electric and magnetic character.
We assume however, that the $\gamma$ decay in the continuum is dominated by dipole transitions and will try to model the $\gamma$-strength function theoretically.
It is commonly adopted that the E1 strength is determined by the giant electric dipole resonance (GEDR) at high $\gamma$ energies.
More doubtful is the assumption that the tail of the GEDR governs the E1 strength at low $\gamma$ energies when the tail approaches zero~\cite{voinov}.
A model~\cite{kad} taking into account the energy and temperature dependence of the GEDR width is often utilized to describe the experimental data~\cite{kop}.
The M1 strength function also plays an important role governing the $\gamma$ emission for low $\gamma$ energies. 
Experiments indicate the existence of an M1 giant resonance due to spin-flip excitations in the nucleus \cite{kop2}. 

\begin{figure}[p]
\centering
\includegraphics[width=17cm]{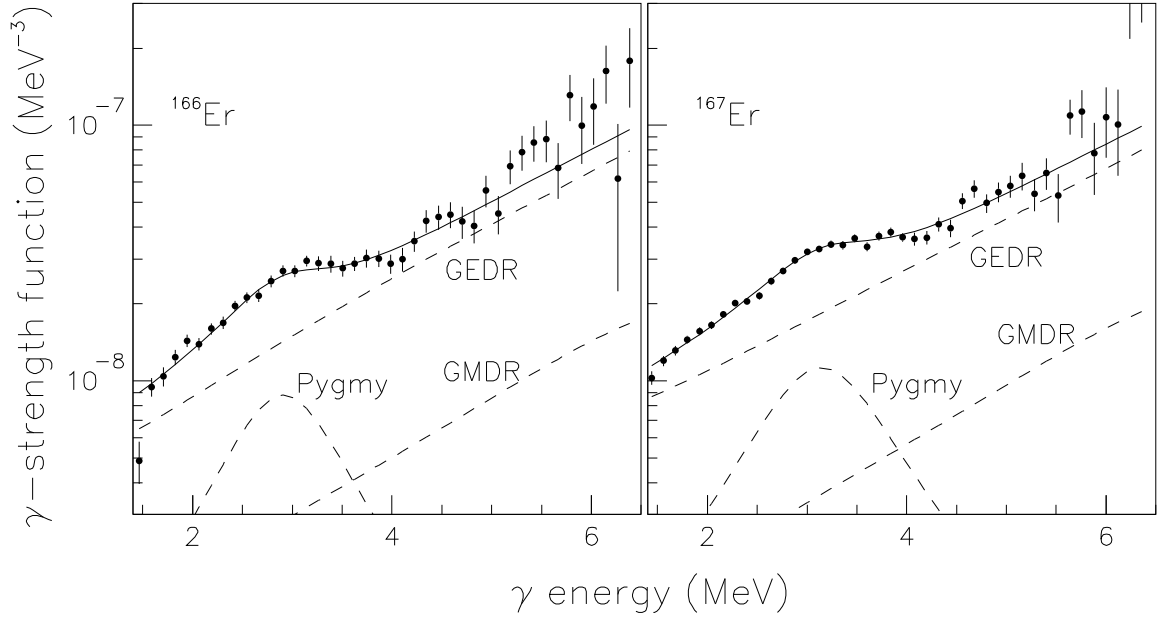}
\caption{The experimental $\gamma$-ray strength function (data points) of $^{166}$Er (left) and $^{167}$Er (right).
The solid line is the fit to the data by the theoretical model.
The dashed lines are the respective contributions of the GEDR, the GMDR, and the pygmy resonance to the total theoretical strength function.}
\label{fig:sffit}
\end{figure}

In Fig.~\ref{fig:sffit}, the experimental $\gamma$-strength function is fitted by a theoretical strength function taking into account both the giant electric dipole resonance and the spin-flip resonance.
In addition, a weaker resonance at lower energies is needed in order to fit the experimental data.
Because of the much lower strength compared to the GEDR, the resonance is denoted the pygmy resonance. 
To account for the E1 radiation, the model~\cite{kad}
\begin{equation} f_{\mathrm{E1}}(E_\gamma)=\frac{1}{3\pi^2\hbar^2c^2} \frac{0.7\sigma_{\mathrm{E1}}\Gamma_{\mathrm{E1}}^2(E_\gamma^2+4\pi^2T^2)} {E_{\mathrm{E1}}(E_\gamma^2-E_{\mathrm{E1}}^2)^2} \end{equation}
is adopted.
The values for the giant electric dipole resonance parameters $\sigma_{\rm E1}$, $\Gamma_{\rm E1}$ and $E_{\rm E1}$ are taken from~\cite{AogN}.
The temperature parameter $T$ is utilized as a constant fit parameter.
The M1 radiation is described by 
\begin{equation}
f_{\mathrm{M1}}(E_\gamma)=\frac{1}{3\pi^2\hbar^2c^2} \frac{\sigma_{\mathrm{M1}}E_\gamma\Gamma_{\mathrm{M1}}^2} {(E_\gamma^2-E_{\mathrm{M1}}^2)^2+E_\gamma^2\Gamma_{\mathrm{M1}}^2}, 
\label{eq:M1}
\end{equation}
where $\sigma_{\rm M1}$, $\Gamma_{\rm M1}$ and $E_{\rm M1}$ are the giant magnetic dipole resonance parameters, which are taken from~\cite{tecdoc}.
The pygmy resonance is here described with a similar Lorentzian function $f_{\rm py}$ as Eq.~(\ref{eq:M1}), where the pygmy-resonance strength $\sigma_{\rm py}$, width $\Gamma_{\rm py}$ and centroid $E_{\rm py}$ have been fitted in order to adjust the total theoretical strength function
\begin{equation}
f=K(f_{{\mathrm{E}}1} + f_{{\mathrm{M}}1})+f_{\mathrm{py}}
\end{equation}
to the experimental data.
The resulting theoretical $\gamma$-strength functions are shown as solid lines in Fig.~\ref{fig:sffit}. 
The obtained values of the fitting parameters for the pygmy resonance, the normalization constant $K$ and the temperature $T$ are given in Table~\ref{tab:fit}.

The low-energetic resonance has its centroid at $E_{\gamma}=2.98(8)$ MeV in $^{166}$Er and at $E_{\gamma}=3.24(7)$ MeV in $^{167}$Er.
In $(n,\gamma)$ reactions a pygmy resonance is observed~\cite{iga} in the energy region $2.5-3.5$ MeV for nuclei of mass 160-176. 
Both the centroid and the width of the resonance is found~\cite{iga} to increase gradually with neutron number.
In $^{166}$Er the centroid, width and strength of the pygmy resonance are somewhat lower than in $^{167}$Er, fitting very well into the systematics of~\cite{iga}.
The enhanced $\gamma$ strength around $3$ MeV in $^{161,162}$Dy and $^{171,172}$Yb~\cite{voinov} also fit into this systematics, indicating that the $\sim 3$ MeV bump after ($^3$He,$\alpha$) and ($^3$He,$^3$He') reactions probably is of the same origin as the pygmy resonance reported in~\cite{iga}.

The pygmy resonance has been explained by the enhancement of the E1 $\gamma$-strength function \cite{iga}. 
The possibility that it on the contrary is of M1 character can not be excluded.
At an excitation energy around $3$ MeV, there is a concentration of orbital M1 strength in the weakly collective scissors mode~\cite{rich}.
The scissors mode was first observed in electron-scattering experiments~\cite{riht}, and is confirmed by the $(\gamma,\gamma ')$ reaction~\cite{pietralla}.
A systematic survey of centroids of observed M1 strength distributions in even-even nuclei in the $A=134-196$ mass region is given in Ref.~\cite{pietralla}, where the scissors mode of $^{166}$Er is found to be centered at $2.961(26)$ MeV, in perfect agreement with the pygmy resonance found here.

Figure~\ref{fig:sffit} also shows the predicted individual contributions from the giant electric dipole resonance $f_{E1}$, the giant magnetic dipole resonance $f_{M1}$ and the pygmy resonance $f_{\rm py}$ to the $\gamma$-strength function.
The strength function is generally dominated by E1 radiation.
The M1-strength function $f_{M1}$ is always $\sim 20$\% lower than the E1-strength function $f_{E1}$.
Moreover, the pygmy resonance contributes considerably around $3$ MeV.

The normalization constant $K$ is close to 1 for $^{167}$Er (Table \ref{tab:fit}), showing that the adopted theoretical model reproduces the absolute values of the $\gamma$-strength functions very well. 
The deviation from $1$ can be explained by uncertain values of the GEDR parameters for the investigated nuclei and uncertainties in the experimental normalization of the $\gamma$-strength functions.  
Also the $K$ value for $^{166}$Er is reasonable, provided that the adopted value for the neutron-resonance radiative width is reliable.

\section{Conclusion}
Levels in $^{166}$Er and $^{167}$Er in the excitation region up to the neutron separation energy were populated with the ($^3$He,$\alpha$$\gamma$) and ($^3$He,$^3$He'$\gamma$) reaction, respectively.
The level density and $\gamma$-ray strength function of $^{166}$Er and $^{167}$Er are determined from their corresponding primary $\gamma$-ray spectra.

Thermodynamic observables are deduced from the level density and display signatures of phase-like transitions within the micro-canonical and the canonical ensemble, interpreted as the transition from a strongly pair-correlated phase to an uncorrelated phase.
Micro-canonical thermodynamics give the possibility of investigating the successive breaking of nucleon pairs in detail, information which is hidden in the canonical approach.
The canonical ensemble on the other hand, reveals the average properties of the pairing transition.
In addition, the canonical ensemble provides an excellent opportunity to study the different mechanisms governing the thermodynamic properties of odd and even systems.
The increase in the heat capacity with temperature is much steeper in $^{166}$Er than in $^{167}$Er, probably due to the tendency of the odd neutron to block higher quasiparticle excitations.
Various estimates of the critical temperature for the pairing-phase transition is performed, giving the result $T_c\sim 0.5$ MeV.

The experimental $\gamma$-strength function is fitted by a theoretical strength function, assuming that the $\gamma$ decay in the continuum is governed by dipole transitions.
The contribution of electric and magnetic dipole radiation to the $\gamma$-strength function is recognized.
A bump is observed in the $\gamma$-strength function at $3.0$ and $3.2$ MeV in the $^{166}$Er and $^{167}$Er, respectively, and is probably of the same origin as the pygmy resonance found in the $(n,\gamma)$ reaction.

\acknowledgements
Financial support from the Norwegian Research Council (NFR) is acknowledged.

\end{multicols}

\newpage
\begin{table}
\caption{Different extracted values for the critical temperature.
} 
\begin{tabular}{lcc}
Extraction method             &$^{166}$Er&$^{167}$Er\\
                              & $T_c$ (MeV)&$T_c$ (MeV)\\
\hline
micro-canonical caloric curve   &          &0.51(4)\\
micro-canonical entropy       &0.53(10)      &0.55(4)\\
canonical heat capacity       &0.47(10)      &0.59(10)\\
canonical constant temperature&0.44(10)      &0.51(9)\\
\end{tabular}
\label{tab:crit}
\end{table}

\begin{table}
\caption{Parameters obtained from the fitting of the $\gamma$-ray strength function.} 
\begin{tabular}{cccccc}
Nucleus&$E_{\mathrm{py}}$&$\sigma_{\mathrm{py}}$&$\Gamma_{\mathrm{py}}$&$T$&K\\
       &(MeV)            &(mb)                  &(MeV)              &(MeV)& \\ 
\hline
$^{166}$Er& 2.98(8)& 0.30(4)    & 1.3(3)    & 0.31(5)  & 1.3(2)  \\
$^{167}$Er& 3.24(7)& 0.43(4)    & 1.7(2)    & 0.36(2)  & 1.27(6)\\
\end{tabular}
\label{tab:fit}
\end{table}
\clearpage

\clearpage
\clearpage

\clearpage

\clearpage

\clearpage

\clearpage

\clearpage

\clearpage

\clearpage

\cleardoublepage

\clearpage

\end{document}